# Faraday-Rotation in Iron Garnet Films Beyond Elemental Substitutions

Miguel Levy[1], Olga V. Borovkova[2], Colin Sheidler[1], Brandon Blasiola[1], Dolendra Karki[1], François Jomard[3], Mikhail A. Kozhaev[2,4], Elena Popova[3], Niels Keller[3], Vladimir I. Belotelov[2,5]

[1]*Physics Department, Michigan Technological University, Houghton, Michigan 49931, USA*

[2]*Russian Quantum Center, Skolkovo, Moscow Region 143025, Russia*

[3]*Groupe d'Etude de la Matière Condensée (GEMaC), CNRS-UVSQ, Université Paris-Saclay, 78035 Versailles, France*

[4]*Prokhorov General Physics Institute RAS, Moscow 119991, Russia*

[5]*Lomonosov Moscow State University, Moscow 119991, Russia*

**Abstract**
In previous decades significant efforts have been devoted to increasing the magneto-optical efficiency of iron garnet materials for the miniaturization of nonreciprocal devices such as isolators and circulators. Elemental substitutions or proper nano-structuring to benefit from optical resonances have been pursued. However, all these approaches still require film thicknesses of at least several tens of microns to deliver useful device applications, and suffer from narrow bandwidths in the case of optical resonance effects. This Letter reports on a newly discovered enhancement of the Faraday Effect observed experimentally in nanoscale bismuth-substituted iron garnet films. It is shown here that this enhancement is not due to elemental substitution or compositional variations. Nor is it due to photon trapping or resonance effects. Comprehensive experimental and theoretical analysis of the Faraday rotation reveals a dramatic seven-fold amplification in the magneto-optic gyrotropy within only 2 nm of the air-surface interface, corresponding to just a couple of atomic monolayers as a result of symmetry-breaking at the air-film interface. This finding opens up an avenue to the application of monolayer magnetic garnets for the control of light.

Rare-earth iron garnet films are technologically important since they are among the main materials used in magneto-photonics, ultrafast magnetism and spintronics [1, 2]. In particular, they are used extensively in optical circuits for the fabrication of nonreciprocal optical devices. Isolators and circulators, operating as optical filters rely on the non-reciprocal functionality of iron garnets to protect laser sources from reflected downstream light that might otherwise enter the cavity and destabilize the signal [2-4].

Magneto-optical light modulators and magnetic sensors utilize iron garnet films for efficient magneto-optical response and outstanding performance [3-5]. On the other hand, iron garnet thin films are being successfully used in magnonics in experiments on the spin-orbit torque transfer and nonlinear spin-waves dynamics [6-9].

Since the development of yttrium iron garnet in the 1950s, subsequent bismuth and cerium substitutions in the dodecahedral sub-lattice of these materials were found to enhance their Faraday rotation response [1, 5, 10]. Such substitutions amplify the spin-orbit splitting of the 3d $Fe^{3+}$ excited states, impacting the refractive index difference between photons of opposite helicities [1]. The Faraday rotation at wavelengths longer than the absorption band is affected as a result. These discoveries impacted the efficiency of nonreciprocal devices, and made it possible to reduce their footprint and hence help drive their miniaturization. However, no other elemental substitutions with similar effect have been found since the 1990s.

The need for on-chip isolator integration for the manufacturing of optical circuits, especially for telecommunication purposes, has led to extensive research on and prototyping of these devices over the years [11-15]. Smart (or engineered/designed) magnetophotonic crystal and magnetoplasmonic structures incorporating bismuth-substituted iron-garnet films were demonstrated [16-19]. However, better and stronger non-reciprocal efficiency than at present is still needed by industry to further miniaturize these systems down to nanometer size.

This Letter reports on newly discovered enhancement effects observed experimentally in ultrathin rare-earth substituted iron garnet films [20-22]. It is shown here that these effects are not due to elemental substitution or compositional variations. Nor are they due to photon trapping or resonance effects. Theoretical modelling presented in this Letter and verified experimentally, shows that surface reflections do impact the magneto-optic response of these films but cannot account for the large Faraday rotation enhancement observed below 50nm-film-thickness.

There are four physical mechanisms that are capable of producing enhancement in the Faraday rotation in magnetic garnets: a) bismuth or cerium substitutions into the dodecahedral sub-lattice of these crystals, b) optical resonances due to reflections, c) surface symmetry-breaking, and d) quantum confinement. This Letter provides strong theoretical and experimental evidence that the observed effects are not due to rare-earth substitutions nor classical electrodynamic interference, thus pointing to surface symmetry breaking effects as the alternative mechanism of significance in the physical response of ultra-thin iron garnet magneto-optics. *Modelling of these experimental* *results is consistent with a surface enlargement in the magneto-optic gyrotropy parameter responsible for the Faraday effect in these films by a factor of seven within only 2nm of the surface.*

Two different sets of liquid-phase-epitaxy (LPE)-grown films with $Bi_{0.8}Gd_{0.2}Lu_2Fe_5O_{12}$, and $Bi_{0.7}Gd_{0.4}Lu_{1.9}Fe_{4.1}Ga_{0.9}O_{12}$ per-formula-unit (pfu) were studied. These films were fabricated on gadolinium gallium garnet (GGG) (100)-oriented substrates. The films for any set of measurements were all taken from the same wafer and wet-etch-thinned-down sequentially in ortho-phosphoric acid, thus avoiding possible composition differences due to growth conditions.

Light from a CW laser source was used to probe the Faraday rotation at 532nm. Measurements normal to the film surface were conducted in a rotating-polarizer configuration, with an accuracy of 0.001° [3]. Faraday-rotation hysteresis loops were recorded for each sample and the Faraday rotation at saturation was chosen to characterize the response of each film. The paramagnetic signal of the GGG substrate was subtracted out from the overall Faraday rotation signal. The experimental setup is described in [20].

Film thickness was measured via ellipsometry, and Faraday rotation per unit length is plotted in Fig. 1 for both sets of samples as a function of thickness. For the ferrimagnetic films thinner than 50 nm the specific Faraday rotation (SFR) grows dramatically with a decrease in film thickness. One can see that the SFR reaches $10°\mu m^{-1}$ in the 20-nm-thick film, almost 2.5 times greater than SFR in the same thick film. For the thin films we observe oscillations due to interference of the emerging waves due to multiple reflections inside the thin film.

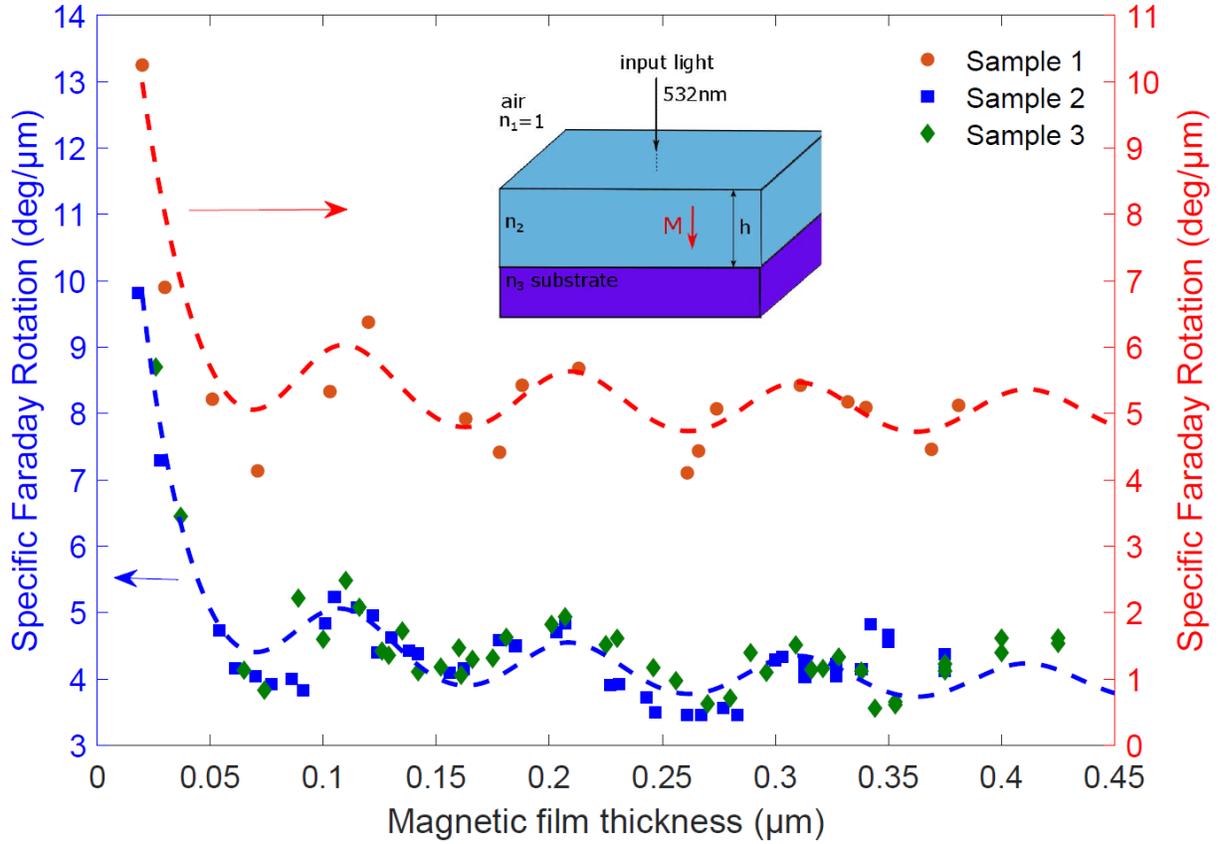

Fig. 1. Specific Faraday rotation at 532 nm versus film thickness for $Bi_{0.8}Gd_{0.2}Lu_2Fe_5O_{12}$ (Sample 1) and $Bi_{0.7}Gd_{0.4}Lu_{1.9}Fe_{4.1}Ga_{0.9}O_{12}$ (Samples 2 and 3), showing a dramatic increase below 50 nm. Inset: Schematic depiction of the samples under consideration.

In this section and in Methods we describe the theoretical model, based on classical electrodynamics, used to analyze the measured Faraday rotation, and the results of this analysis. Linearly polarized light impinges on the magnetic film at normal incidence. The magnetization is directed along the propagation direction, perpendicular to the film surface. Refractive indices and film thickness are as shown in the inset to Fig. 1. The magneto-optic gyration in the ferrimagnetic film is initially assumed to be uniform inside the film and parameterized by the gyrotropy parameter $g$. Multiple reflections are included in the analysis, giving rise to an oscillatory behavior in FR as a function of film thickness $h$.

The exact expression for Faraday rotation is derived in Methods [see Eq. (M3)]. It has two asymptotes, for $h \ll \frac{2\pi}{\lambda}$, ultra-thin films, and for $h \gg \frac{2\pi}{\lambda}$, bulk ferrimagnetic material, with specific Faraday rotations (SFR) given by

$$SFR = \frac{\theta}{h} = \begin{cases} -\frac{g}{2n_2}\frac{2\pi}{\lambda}, & \text{if } h \gg \frac{2\pi}{\lambda} \\ -\frac{g}{1+n_3}\frac{2\pi}{\lambda}, & \text{if } h \ll \frac{2\pi}{\lambda} \end{cases} \quad (1)$$

From (1), the ratio of the SFR at small $h$ with respect to the SFR at large $h$ is, thus,

$$\frac{SFR_{h \to 0}}{SFR_{h \to \infty}} = \frac{2n_2}{1+n_3} \quad (2)$$

The refractive indices at wavelength λ = 532nm are given by $n_3 = 1.980 + 0.003i$ in the GGG substrate, and $n_2 = 2.610 + 0.056i$ in the Bi-substituted iron garnet films for all samples. Therefore, at λ = 532nm this ratio is predicted to be about 1.75, but the experimentally measured ratios are always larger than 2. In particular, $\frac{SFR_{h \to 0}}{SFR_{h \to \infty}} = \frac{10.25}{5} \approx 2.05$, for Sample 1, and $\frac{SFR_{h \to 0}}{SFR_{h \to \infty}} = \frac{9.818}{4.226} \approx 2.32$, for Samples 2 and 3. Thus, the analytical expression derived from classical electrodynamic cannot account for all the effects operating in this system.

Besides the preceding analysis we also performed an electromagnetic modelling based on a rigorous coupled-wave analysis (RCWA) [23, 24]. The result of the corresponding numerical simulation is given in Fig. 2 by the red line (Model 1). One can see that this model does not reproduce the pronounced increase in the SFR obtained experimentally. Therefore, the assumption of constant gyration inside the ferrimagnetic film does not describe the observed abrupt growth in the SFR in ultrathin ferromagnetic films shown in Fig. 1.

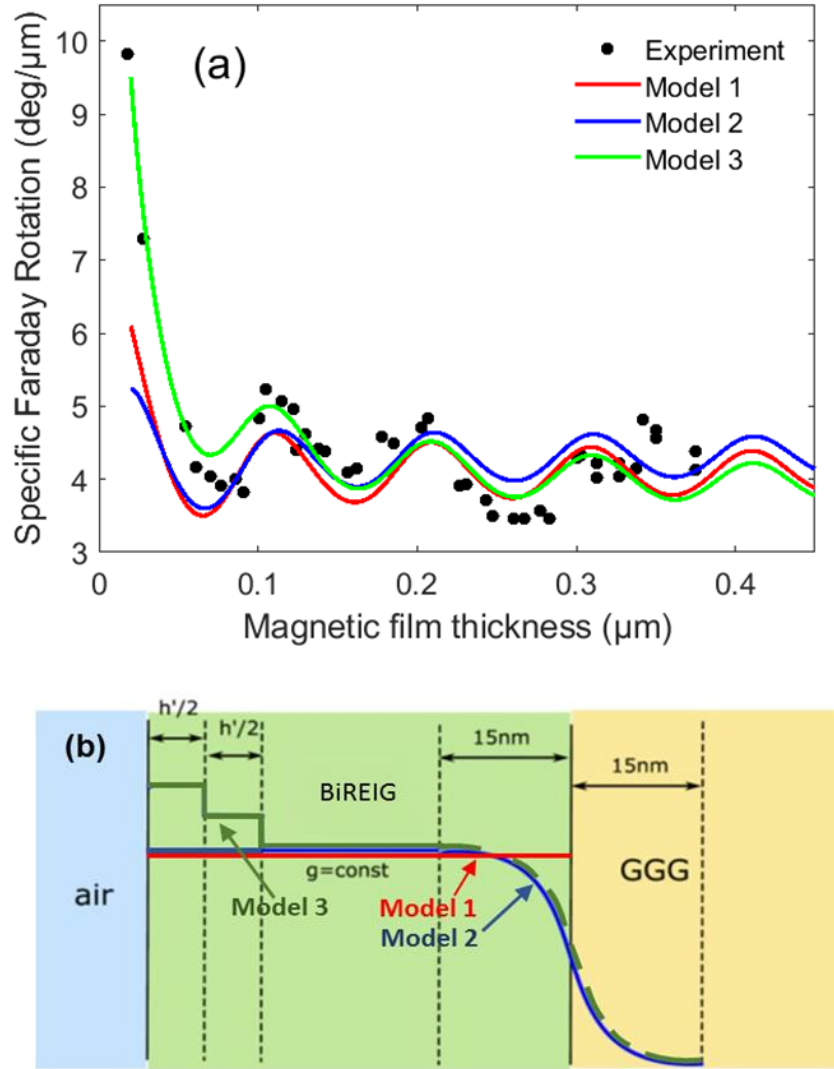

Fig. 2. (a) Theoretical fits to Sample-2 data based on uniform-gyrotropy-parameter in the film (Model 1), gyrotropy-parameter proportional to Bi content in the film (Model 2), and gyrotropy-parameter-enhancement at the top surface (Model 3). A seven-fold *g*-value magnification is predicted within 2 nm of the surface (Model 3), and four-fold magnification over the next 2 nm, as compared to bulk *g*-value. (b) Schematic depiction of the gyrotropy parameter distribution across the film in the three models. BiREIG stands for Bi-rare-earth-substituted iron garnet.

This abrupt growth in SFR in ultrathin films might possibly have been explained by composition gradients in the film/substrate interface, such as an increase in Bi content [1, 5, 10]. However, detailed cross-sectional composition analysis, discussed below, shows that there is no such increase in Bi content, and that the only composition gradient is manifested as a decrease in Bi, Fe, and Lu, content in the transient layer (Fig. 3). The cross-sectional film composition was measured by two different methods, via secondary-ion-mass-spectroscopy (SIMS), Fig. 3(a) and cross-sectional scanning transmission electron microscopy (S-TEM), Fig. 3(b) (Methods).

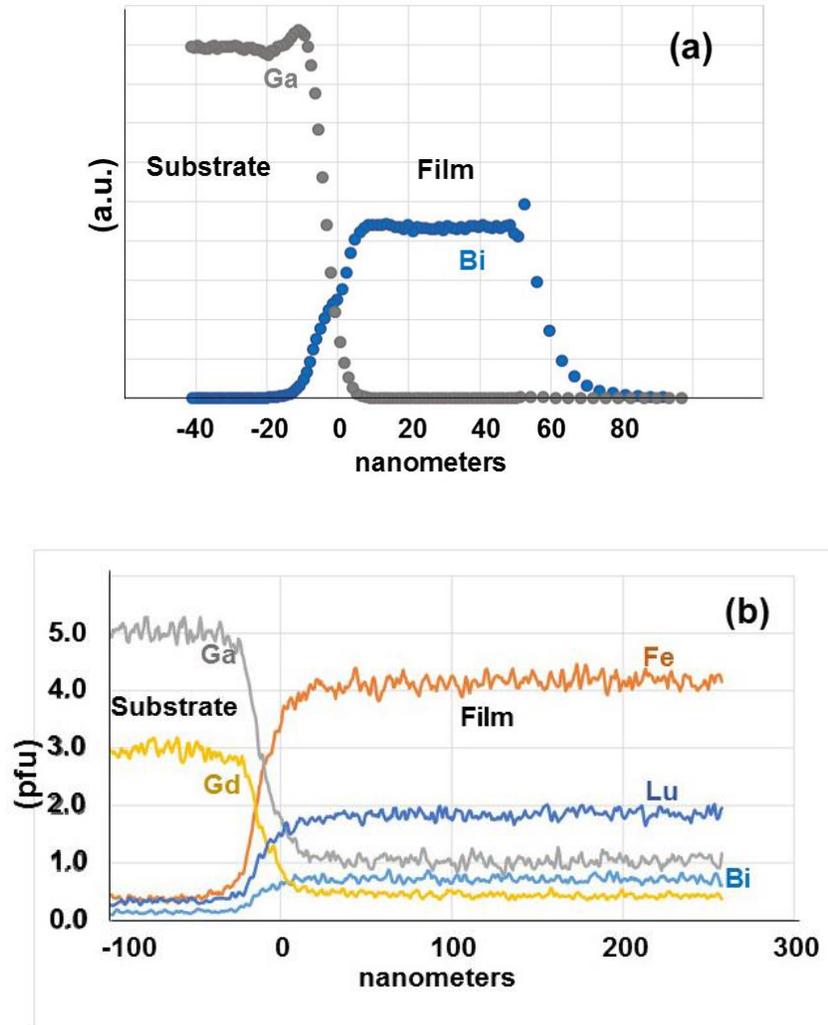

Fig. 3. (a) Bi and Ga concentrations in $Bi_{0.8}Gd_{0.2}Lu_{2.0}Fe_{4.5}O_{12}$ (Sample 1), arbitrary units, determined by SIMS on a 60nm-thick film. Bi concentration is shown multiplied by factor 10 for the sake of clarity. (b) Elemental concentrations (pfu) for Bi, Lu, Gd, Ga and Fe in $Bi_{0.7}Gd_{0.4}Lu_{1.9}Fe_{4.1}Ga_{0.9}O_{12}$ (Samples 2 and 3) determined by cross-sectional S-TEM EDX. Both show uniform concentrations as a function of position above (film) and below (substrate) the transient layer. The zero of horizontal coordinates is set at the film-substrate interface.

Assuming a gyration proportional to the Bi-content, comprising also its decline in the transient layer (Fig. 2(b), blue line), the numerical simulation produces an even poorer coincidence with the experimental data than Model 1, constant gyration. This is shown by the blue line (Model 2) in Fig. 2(a). Notice that Model 1 in effect already assumes a significant increase in magneto-optic gyration in the interfacial region, if one considers the decrease in Bi content in the transient layer. Yet the calculated SFR in that case still remains well below the experimental data.

To achieve coincidence with the experimental data, we allow the gyrotropy parameter $g$ to experience an increase near the air/film interface. Best fits to the experimental data are produced by a rise in $g$ value taking place over a very thin layer adjacent to the surface, which is h′ = 4nm thick. In Model 3 the gyration evinces a near-surface step-like growth, as depicted in Fig. 2(b). One can see from the Fig. 2 (Model 3) that this assumption provides very good agreement with the SFR, including for sub-50nm-thick ferrimagnetic films. These fits predict that the magneto-optic gyrotropy parameter exhibits a seven-fold amplification in magnitude within 2 nm of the surface, and a four-fold amplification in the next 2 nm, over its bulk value (Methods).

To summarise, detailed Faraday rotation data and electrodynamic analysis of ultra-thin Bi-substituted iron garnet films predict a several-fold increase in near-surface magneto-optic gyrotropy within 2 nm of the surface. Electron dispersive X-ray spectroscopy and SIMS confirm that the measured Faraday rotation enhancement in ultra-thin films is not due to compositional non-uniformities near the film-substrate interface, thus pointing to surface symmetry breaking effects as the origin of these results. This dramatic enhancement in the gyrotropy parameter occurs within a few monolayers of the films, as the unit-cell in these iron garnets is ~1.2 nm. These results hold a significant potential for the nanofabrication of nonreciprocal devices, given the technological role of these materials in optical isolator and circulator components.

**Acknowledgements**

This study was supported by Russian Foundation for Basic Research (project no. 16-32-60135 mol_a_dk), and by Foundation for the advancement of theoretical physics BASIS. ML and DK acknowledge support from the Henes Center for Quantum Phenomena. The authors also thank Ashim Chakravarty for valuable discussions and Pinaki Mukherjee for assistance with the cross-sectional EDX S-TEM measurements of film composition.

## Methods

### Compositional analysis

SIMS measurements were performed with $O_2^+$ bombardment and positive ion detection mode using an IMS-7f (CAMECA) micro-analyzer. The primary beam was rastered over an area of 125 x 125 µm$^2$ and the secondary ions were collected from the central part of this area (diameter 33 µm). In order to avoid charging effect, the sample was coated with a layer of gold (50 nm) and electron flooding was employed using a normal incidence e-gun.

Electron micrographs and Energy dispersive X-ray spectroscopy (EDX) maps were obtained on $Bi_{0.7}Gd_{0.4}Lu_{1.9}Fe_{4.1}Ga_{0.9}O_{12}$ (Samples 2 and 3) using an FEI Titan Themis aberration-corrected S-TEM operated at 200kV. The point resolution in this aberration-corrected mode is 0.08nm. The microscope is fitted with a Super-XTM X-ray detector, which is a combination of 4 detectors for fast X-ray mapping in S-TEM mode. For the present experiment, one-nm resolution EDX maps were taken with an average beam current of 100pA. The size of the maps were 512 X 512 pixels and 50µs/pixel dwell time was used for collecting the signal. All maps are generated by summing over 10 frames. Drift correction during data collection and subsequent analysis were performed using Velox software. These results are shown in Fig. 4(a), with a standard deviation of ±0.05 pfu for Bi, and ±0.11 pfu for Fe.

### Magnetic samples

Two different sets of liquid-phase-epitaxy (LPE)-grown films with $Bi_{0.8}Gd_{0.2}Lu_2Fe_5O_{12}$, and $Bi_{0.7}Gd_{0.4}Lu_{1.9}Fe_{4.1}Ga_{0.9}O_{12}$ per-formula-unit (pfu) were studied. These films were grown on (100) gadolinium gallium garnet (GGG) substrates. The bulk gyration, obtained from measurements in $2\mu m-$thick films, is different for the two different composition. For $Bi_{0.7}Gd_{0.4}Lu_{1.9}Fe_{4.1}Ga_{0.9}O_{12}$ pfu films (Samples 2 and 3), the bulk $g = 0.029 + 0.003i$, and for $Bi_{0.8}Gd_{0.2}Lu_2Fe_5O_{12}$ (Sample 1), it is $g = 0.038 + 0.004i$.

### Theoretical analysis

The transmission coefficients for right- and left-circularly polarized light are given by

$$t^\sigma = \left(\frac{E_3^\sigma}{E_i^\sigma}\right) = \frac{2}{(1+n_3)\cosh(ik_0\tilde{n}_2 h) - \left(\frac{n_3}{\tilde{n}_2} + \tilde{n}_2\right)\sinh(ik_0\tilde{n}_2 h)}, \quad \text{(M1)}$$

where $\sigma = \pm 1$ denotes the optical polarization, $E_i^\sigma$, $E_3^\sigma$ are the amplitudes of the incident and transmitted to the substrate light circularly polarized components, respectively, $n_3$ is the refractive index of the substrate, and where the refractive index of the ferrimagnetic film can be written as $\tilde{n}_2 = \sqrt{n_2^2 \pm g}$, $k_0 = \frac{2\pi}{\lambda}$ is the wave number in vacuum, and $\lambda$ is the wavelength [25].

From Eq. (M1) and the fact that $\sinh(ix) = i\sin x$ and $\cosh(ix) = \cos x$, one can find an expression for the transmitted light in terms of the component of the incident light

$$E_3^\sigma = \frac{2E_i^\sigma}{(1+n_3)\cos\left(\frac{2\pi}{\lambda}h\sqrt{n_2^2 \pm g}\right) - i\left(\frac{n_3}{\sqrt{n_2^2 \pm g}} + \sqrt{n_2^2 \pm g}\right)\sin\left(\frac{2\pi}{\lambda}h\sqrt{n_2^2 \pm g}\right)}. \quad \text{(M2)}$$

Consequently, the Faraday rotation (FR), taking into account surface reflections and absorption losses, can be expressed as

$$\theta = \frac{1}{2}\arg\left(\frac{E_3^+}{E_3^-}\right) =$$

$$= \frac{1}{2}\arg\left[\frac{(1+n_3)\cos(k_0\sqrt{\varepsilon_2 - g}\,h) - i\left(\frac{n_3}{\sqrt{\varepsilon_2 - g}} + \sqrt{\varepsilon_2 - g}\right)\sin(k_0\sqrt{\varepsilon_2 - g}\,h)}{(1+n_3)\cos(k_0\sqrt{\varepsilon_2 + g}\,h) - i\left(\frac{n_3}{\sqrt{\varepsilon_2 + g}} + \sqrt{\varepsilon_2 + g}\right)\sin(k_0\sqrt{\varepsilon_2 + g}\,h)}\right], \quad \text{(M3)}$$

where $\varepsilon_2 = n_2^2$.